\documentclass[conference]{IEEEtran}
\IEEEoverridecommandlockouts
\usepackage{cite}
\usepackage{amsmath,amssymb,amsfonts}
\usepackage{algorithmic}
\usepackage{graphicx}
\usepackage{textcomp}
\usepackage[numbers]{natbib}
\usepackage{url}
\usepackage{hyperref}
\usepackage{xcolor}

\usepackage{amssymb, amsthm, amsmath}
\usepackage{color}
\usepackage[ruled,vlined]{algorithm2e}
\usepackage{subcaption}

\newcommand{\R}{\mathbb{R}}

\newcommand{\cD}{\mathcal{D}}

\newcommand{\cN}{\mathcal{N}}
\newcommand{\cZ}{\mathcal{Z}}
\newcommand{\cT}{\mathcal{T}}
\newcommand{\cX}{\mathcal{X}}
\newcommand{\cW}{\mathcal{W}}

\newcommand{\la}{\lambda}
\newcommand{\tla}{\tilde{\lambda}}

\newcommand{\bv}{\mathbf{v}}
\newcommand{\tbv}{\tilde{\mathbf{v}}}

\newcommand{\bu}{\mathbf{u}}
\newcommand{\bw}{\mathbf{w}}
\newcommand{\bd}{\mathbf{d}}

\SetKwComment{Comment}{/* }{ */}
\newcommand{\ignore}[1]{{}}
\newtheorem{theorem}{Theorem}[section]

\def\BibTeX{{\rm B\kern-.05em{\sc i\kern-.025em b}\kern-.08em
    T\kern-.1667em\lower.7ex\hbox{E}\kern-.125emX}}

\begin{document}

\title{Parallel Computation of Multi-Slice Clustering of
Third-Order Tensors\\
}

\makeatletter 
\newcommand{\linebreakand}{%
  \end{@IEEEauthorhalign}
  \hfill\mbox{}\par
  \mbox{}\hfill\begin{@IEEEauthorhalign}
}
\makeatother 

\author{\IEEEauthorblockN{Dina Faneva Andriantsiory}
\IEEEauthorblockA{\textit{Laboratoire d'Informatique de Paris Nord} \\
\textit{LIPN - UMR CNRS 7030}\\
Villetaneuse, France \\
andriantsiory@lipn.univ-paris13.fr}
\and
\IEEEauthorblockN{Camille Coti}
\IEEEauthorblockA{\textit{École de Technologie Supérieure} \\
Montreal, Canada \\
camille.coti@etsmtl.ca}
\and
\linebreakand
\IEEEauthorblockN{Joseph Ben Geloun}
\IEEEauthorblockA{\textit{Laboratoire d'Informatique de Paris Nord} \\
\textit{LIPN - UMR CNRS 7030}\\
Villetaneuse, France \\
bengeloun@lipn.univ-paris13.fr}
\and
\IEEEauthorblockN{Mustapha Lebbah}
\IEEEauthorblockA{\textit{DAVID Lab} \\
\textit{University of Versailles, Université Paris-Saclay}\\
Versailles, France \\
mustapha.lebbah@uvsq.fr}
}

\maketitle

\begin{abstract}
Machine Learning approaches like clustering methods deal with massive datasets that present an increasing challenge. We devise parallel algorithms to compute the Multi-Slice Clustering (MSC) for 3rd-order tensors. The MSC method is based on spectral analysis of the tensor slices and works independently on each tensor mode. Such features fit well in the parallel paradigm via a distributed memory system. We show that our parallel scheme outperforms sequential computing and allows for the scalability of the MSC method. 
\end{abstract}


\section{Introduction}
Tensorial or multidimensional data are prominent in machine learning (ML) and arise in sundry contexts such as neuroscience, see \cite{dataneuroscience}, genomics data, see  \cite{datagenomics1} \cite{datagenomics2}, computer vision, for instance  \cite{datacomputerVision}, and several other domains. As a privileged ML approach, the clustering analysis extracts patterns in data and the massive nature of these datasets and their high-dimensionality require efficient algorithms. High performance computing is an natural approach to take advantage of considerable computing resources and handle larger datasets to reach exploitable results.

Addressing on 3D datasets, in other words 3rd-order tensors, 
various methods offers tools for clustering such multidimensional data. Some procedures demand the number of clusters as an input (this is the case of Tucker or CP+k-means, Multiway Clustering via Tensor Block Models (TBM) \cite{WangMultiwayClustering2019},  Heterogeneous Tensor Decomposition for Clustering via Manifold Optimization  \cite{Boumal2014manifoldOptimisation}, the Tensor Latent Block Model for Co-Clustering \cite{RafikaB2020TLBM}). Other clustering methods focus on the cluster extraction of given size (the Tensor Bi and Triclustering require the desired cluster size \cite{FeiziNIPS2017}).
Whether it is the number of clusters or their size, these parameters are difficult to estimate on real life data. Hence, due to the lack of ground truth for validation,  using and the performance comparison of such algorithms with others stay questioned \cite{helfmann2018hyperparameter}. The quest for other universal methods without these drawbacks continues.

The Multi-Slice Clustering (MSC) defines a method that addresses the previous issue for 3rd-order tensor dataset  \cite{andriantsiory2021multislice}. MSC uses a threshold parameter which gauges the similarity in the delivered cluster. It has been recently extended to tensor datasets bearing numerous clusters \cite{andriantsiory2023dbscan} using the DBSCAN method \cite{ester1996densitydbscan} and further generalized to multiway clustering \cite{andriantsiory2023multiway}. These ranges of properties have motivated our work to devise high performance computing for such a clustering approach. 

The advent of new ML algorithms  in recent years has obviously stimulated the use of parallel and distributed computing with a view to coupling both relevance and performance, see for instance  \cite{park2002distributed, Tsoumakas2009, Bacarella2013,Bendechache2018, Lu_2020}. In 2020, Lu et al introduced a density-based improved k-means distributed clustering algorithm using the Spark parallel framework \cite{Lu_2020}. If this proposal 
improves one of the most emblematic clustering method, namely k-means, this algorithm has no extension to higher dimensional datasets. As far as multidimensional data processing is concerned, a few approaches have come to light. Among these works, it is worth advertising Bendechache's proposal for tensorial datasets \cite{Bendechache2018}:  
it builds a dynamic programming approach to speed up the response and deliver accurate results for clustering analysis. 
We will present a new method implementing parallel 
computation for tensor dataset clustering based on 
the feature that some clustering algorithms for multidimensional
data work dimension by dimension.

In this paper, we introduce a parallel version of the MSC algorithm to take advantage of parallel, distributed  memory systems and scale on computing resources. Our goal is to preserve the clustering quality with respect to the sequential version, and to take advantage of high performance computing facilities. Opting for a different strategy from previous works, MSC computes dimensions one after another, and its parallelized version relies on this property. Note that, during the clustering, the tensor dataset is read in a given dimension, decomposed into slices (a slice being a matrix within the tensor at a fixed tensor index) and a spectral analysis applies to each slice independently from the other ones. The computation  distributes between multiple processes, and each process holds only the necessary data for its computation. A performance evaluation of our algorithm shows that it successfully scales with the number of processes. 

The outline of this paper is as follows: section \ref{sec:mode1} reviews the MSC method. 
Section \ref{parallel} presents a parallel MSC algorithm. 
First, we present our multi-level parallel algorithm, then explain the data distribution 
and layout between processes, and detail the local computation 
performed by local processes. 
 Finally, we collect the partial results to build a clustering. 
Section \ref{Parall-experiments} presents a performance evaluation of the parallel MSC and discusses the observed results. 
Short comments on prospects close the paper.

\section{MSC for 3rd-order tensors: an overview} \label{sec:mode1}

Let $m_1,m_2$ and $m_3$ be
three positive integers. 
A 3rd-order tensor $\cT\in\R^{m_1\times m_2\times m_3}$  is a 3 dimensional array, each dimension of which is called {\bf mode} of the tensor (see figure \ref{fig:tensor}).
	\begin{figure}[h!]
	\centering

		\includegraphics[width=0.3\textwidth]{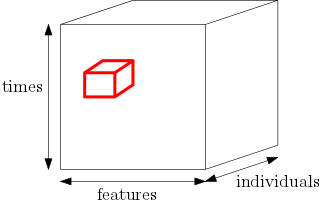}

	\caption{ Representation of a 
	3-order tensor data set and a tricluster highlighted.
 }
	\label{fig:tensor}
\end{figure}

$\cT$ is modeled  as $\cT = \cX + \cZ$, where $\cX$ is the signal tensor and $\cZ$ is the noise tensor, i.e. its entries are i.i.d. random variables obeying the standard normal law. We assume that $\cX$ is decomposed in a rank-1 tensor: 
\begin{equation}
\cT = \cX + \cZ = \gamma \bw \otimes \bu \otimes\bv  + \cZ, \label{eq:problem}
\end{equation}
where $\bw \in \R^{m_1},  \bu \in \R^{m_2}, \bv \in \R^{m_3}, \gamma >0$.

Given a mode $i$ of $\cT$, for $j_i \in [m_i]$, we call $j_i$-th {\bf slice} 
of mode $i$ of the tensor $\cT$, 
the set of entries of $\cT$ obtained by fixing $j_i$ in its $i$th mode. This set of entries 
is obviously structured as a matrix. 

The MSC solves a triclustering problem. Assume that $J_i$ is index's set of the cluster in  the $i$-th mode. Then, 
the entries of the tensor defined by $J_1\times J_2\times J_3$ represent a tricluster   (figure \ref{fig:tensor})  if these entries are similar in a well-defined sense. 
	The MSC   aims at determining and gathering the indices of the matrix slices that are similar 
 (a definition will follow); it operates in each tensor dimension and combines the results (figure \ref{fig:tensor_to_msc}). 

	\begin{figure}[htbp]
	 		\centering
	\includegraphics[width=8.4cm, height=2.3cm ]{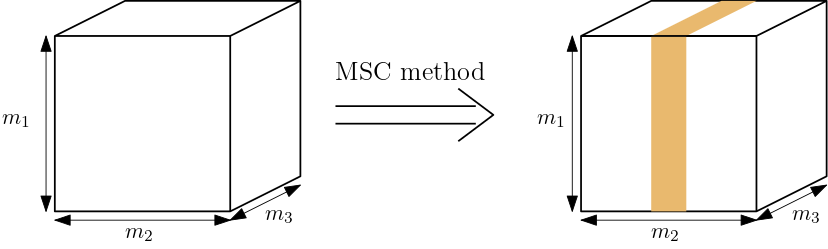}
	 	\caption{Illustration of the output of MSC in one mode.}
   			\label{fig:tensor_to_msc}
	 \end{figure}
 In the following, we focus on mode-1 of the tensor, the reasoning readily extends to other modes. 
 We introduce (in Matlab notation)
$T_i = \cT(i,:,:)=X_i + Z_i,$ $i\in[m_1]$  the $i$-th slice in the mode-$1$.  Since we have a signal tensor of rank-1, the covariance matrix
$C_i= T_i^tT_i $ is decomposed via spectral analysis to give  
\begin{eqnarray}
C_i & =& \left\{ 
\begin{array}{l l}
  \la_i \bv_i\bv_i^t + W_i & \quad \text{if } i\in J_1\\
  Z_i^t Z_i & \quad \text{otherwise}\\ \end{array} \right. \label{covvector} 
\end{eqnarray}
where $W_i = X_i^t Z_i + Z_i^t X_i + Z_i^t Z_i$ and $\la_i = \gamma_i^2 (w(i))^2$
stands for the signal strength. 
Targeting the noise slice, 
each row of $Z_i$ is real an independently drawn form $\cN_{m_3}(0,I)$, the $m_3$-variate normal distribution with zero mean and covariance matrix $I$. The matrix $Z_i^t Z_i$ has a white Wishart distribution $\cW_{m_3}(m_2, I)$,
see \cite{johnstone2001}. Its largest eigenvalue distribution $\la$  (when $m_2\rightarrow \infty, m_3\rightarrow \infty$ and $\frac{m_2}{m_3}\rightarrow \gamma>1$) obeys the limit in the distribution sense
	\begin{equation}\label{convergenceDistribution}
		\frac{\la_i - \mu_{m_2m_3,1}}{\sigma_{m_2m_3,1}}\xrightarrow[]{\cD} F_1
	\end{equation}
	where $F_1$ is the Tracy-Widom (TW) cumulative distribution function (cdf), and
\begin{eqnarray}
	\mu_{m_2m_3,1} &=& (\sqrt{m_2-1} + \sqrt{m_3})^2\crcr
	\sigma_{m_2m_3,1} &=& \sqrt{\mu_{m_2m_3,1}} \big(\frac{1}{\sqrt{m_2 - 1}}+\frac{1}{\sqrt{m_3}}\big)^{\frac{1}{3}}.
\end{eqnarray}
 Although \eqref{convergenceDistribution} is true in the limit, it is also shown that it provides a satisfactory approximation from $m_2, m_3 \ge 10$.
 
	 Let $\underline\la_i$ be 	the top eigenvalue 	and $\tbv_i$ be the  	top eigenvector of $C_i$.
	We construct the matrix 
	$V = \begin{bmatrix}
	\tla_1\tbv_1 & \cdots & \tla_{m_1}\tbv_{m_1}	\end{bmatrix}$ 
	where  we set $\tla_i$ to $ \underline\la_i/\la, \forall i\in[m_1]$, and $\la =  \max(\underline\la_1,\cdots,\underline\la_{m_1})$.
	Let ${\bf C} = (c_{ij})_{i,j\in[m_1]}$ be the matrix with each entry defined by 
$		c_{ij}  = | \langle \tla_i \tbv_i,\tla_j \tbv_j \rangle| = \tla_i\tla_j| \langle  \tbv_i, \tbv_j \rangle| \le 1$. 

We are in a position to define the  notion of similarity between matrix 
slices \cite{andriantsiory2021multislice} (see Definition III.2): we call the $i$-th and $j$-th slices $\epsilon$-similar, if for a small $\epsilon>0$ we have,
$c_{ij} \geq 1 - \frac{\epsilon}{2}$.

Let $\bd$ denote the marginal sum vector of the matrix ${\bf C}$. Each entry of $\bd$ is defined by: 
	$ d_i =  \sum_{j\in [m_1]} c_{ij} $. 
The following statement holds \cite{andriantsiory2021multislice}: 
\begin{theorem}\label{thm1}	Let $l = |J_1|$, assume that $\sqrt{\epsilon}\le \frac{1}{m_1-l}$. $\forall i,n\in J_1 $, for $\la = \Omega(\mu)$, there is a constant $c_1>0$ such that
\begin{equation}
	\label{didn}
		|d_i - d_{n}| \le l\frac{\epsilon}{2} + \sqrt{\log(m_1-l)}
	\end{equation}
	holds with probability at least $1-e(m_1-l)^{-c_1}$.
\end{theorem}

The MSC algorithm derives from this theorem.

{\small 
\begin{algorithm}[h]
		\SetAlgoLined
		\KwResult{sets $J_1, J_2$ and $J_3$}
		\KwIn{$3$-order tensor $\cT\in\R^{m_1\times m_2\times m_3}$, real parameter $\epsilon>0$}
		\For{$j$ in $\{1,2,3\}$}{
			Initialize the matrix $M$\;
			Initialize $ \la_0\gets 0$\;
			\For{ i in $\{1,2,\cdots,m_j\}$}{
				Compute : $C_i\gets T_i^t T_i $ \;
				Compute the top eigenvalue and eigenvector $(\la_i, \bv_i)$ of $C_i$\;
				Compute $M(:,i) \gets \la_i * \bv_i$\;
				\If{$\la_i >\la_0$}{
					$\la_0 \gets \la_i$ \;
				}
			}
			Compute : $V \gets M / \lambda_0$\;
			Compute : ${\bf C } \gets |V^t V|$ \;
			Compute : $\mathbf{d}$ the vector marginal sum of ${\bf C }$ and sort it\;
			Initialize the elements of $J_j$ using the max gap in $\mathbf{d}$\;
			Compute : $l\gets |J_j|$\;
			\Repeat{The convergence of the elements of $J_j$(theorem \ref{thm1})}{			
			Update the element of $J_j$ (excluding $i$ s.t. $d_i$ is the smallest value that violates theorem \ref{thm1})\; 
				Compute $l$\;		
			}
		}
		\caption{Multi-slice triclustering}
		\label{algo:multi-slice}
	\end{algorithm}
	}

 \section{Parallel algorithm}
\label{parallel}

This section details the parallel MSC algorithm derived from algorithm \ref{algo:multi-slice}. 

\subsection{Multi-level parallel algorithm}
From algorithm \ref{algo:multi-slice}, we can easily observe that the 
determination of the 
clusters in modes $j=1,2$ and $3$ are
independent of each other, i.e. there is no data dependency between them during their computation. Thus, the cluster resolution of each tensor mode   (the loop in $j$) can be computed in parallel on multiple processes with no memory shared between processes. We create three disjoint groups of processes such that each group computes one mode of the tensor. 

A second crucial remark is that the computation of the top eigenvalue and eigenvector of the different slices are, likewise, independent from each other. This task 
 can be performed in parallel by different processes.
Hence, we partition the slices between the processes of each group. 
All the processes within a given group participate in that computation. In the next stage, the results are combined to extract the maximal eigenvalue of all slices that will be used to normalize all the vectors.

\begin{algorithm}
		\SetAlgoLined
  	\KwResult{sets $J_1, J_2$ and $J_3$}
		\KwIn{$3$-order tensor $\cT\in\R^{m_1\times m_2\times m_3}$, real parameter $\epsilon>0$}
        \KwIn{$global\_rank$: the process's rank on the global communicator}
        \KwIn{$group\_communicator$,$intercommunicator$: communicators }

        \Comment{The input data is distributed such in a way that each process has a part of the slice data}
        
        \Comment{One group of processes aims to find the cluster in a given mode}
        \Comment{Computation of the matrix $M$ of top eigenvectors}
        \For {Each slice $i$}{
            Compute : $C_i\gets T_i^t T_i $ \;
		   Compute the top eigenvalue and eigenvector $(\la_i, \bv_i)$ of $C_i$\;
            Build the sub-columns of $M$\;
            Compute the local maximum of $\lambda$\;
        }
        
        \Comment{Build the full matrix of top eigenvectors}
        
         MPI\_Allgatherv(eigenvectors, $group\_communicator$) \;
   MPI\_Allgatherv($M$, $group\_communicator$) \; 
        \Comment{Maximum global of $\lambda$ for each group}
        $\lambda_{\max} \gets$ MPI\_Allreduce($\lambda$, $MAX$, $group\_communicator$) \;
        \Comment{Normalization}
        $V\_local \gets M/ \lambda_{max}$\;

        \Comment{Similarity}
        Compute the local part of the matrix $C$\;
        Compute the local part of entry of $\mathbf{d}$\;
        
        \Comment{Build the full vector $\mathbf{d}$}
        MPI\_Gatherv( $root\_process$, $\mathbf{d}$, $intercommunicator$ ) \;
        
        \If{ $global\_rank == root\_process$ }{
			Initialize the elements of $J$ using the max gap in $\mathbf{d}$\;
			Compute : $l\gets |J|$\;
			\Repeat{The convergence of the elements of $J$ (theorem \ref{thm1})}{			
			Update the element of $J$ (excluding $i$ s.t. $d_i$ is the smallest value that violates theorem \ref{thm1})\; 
				Compute $l$\;		
			}
       }
    	\Comment{Gather the element of the clusters from each mode}
        MPI\_Gatherv( $J$, $root\_process$, $intercommunicator$) \;
		\caption{Parallel multi-slice triclustering}
		\label{algo:multi-slice-parallel}
  
	\end{algorithm}

In practice, we chose to implement this algorithm using the Message-Passing Interface (MPI). MPI is the \emph{de facto} standard for programming parallel applications over shared memory systems. It features one-sided and two-sided peer-to-peer communication routines, collective communication routines, and dynamic process management utilities. The implementation of our algorithm using the MPI interface is straightforward, using two-sided communications and collective routines to gather the values computed for each mode, and group communicators to manage the sets of processes easily.

\subsection{Data distribution and distributed layout}

The processes ($p$ in total) are divided into groups of $p/3$ processes. We denote by group $j$ the group of processes aiming to compute the $J_j$ along the mode-$j$ of the tensor. The computation workload on each mode is expected to be the same, ensuring load balancing between the groups of processes. 

The groups of processes are working on the same data, but each group works on it in a different direction. Hence, the data is available in every process group but distributed between the processes of each group in a different direction. This distribution is represented in Figure \ref{fig:decomposition}: the first group of processes computes mode-1, so the slices are distributed between the processes of this group along this mode (solid color fill); the second group of processes computes mode-2, so the slices are distributed between the processes of this group along this mode (dotted color fill); the third group of processes computes mode-3, so the slices are distributed between the processes of this group along this mode (sketched color fill). The slices are distributed evenly or almost evenly between the processes, and the computation of each slice is expected to take the same time as the other one, hence, here too, the computation is balanced between the processes of each group.

\begin{figure}
    \centering
    \includegraphics[width=\linewidth]{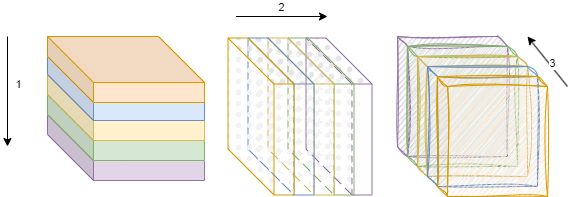}
    \caption{Data distribution between the processes computing each mode.}
    \label{fig:decomposition}
\end{figure}

This data distribution approach differs from the 3D FFT because, with MSC, each mode is independent of the other ones. With the 3D FFT, the result of the computation of the first direction is needed by the computation of the second direction \cite{takahashi2010implementation}. Therefore, with the FFT, the three directions cannot be computed in parallel and the matrix needs to be rotated to reset the data distribution. Again, with MSC, the three modes can be computed independently, so each group which is assigned a mode distributes the data along the corresponding direction.

\subsection{Local computation performed by each process}

In this section, we focus on the computation of the first mode ($j=1$ in algorithm \ref{algo:multi-slice}), which aims to determine the element of the cluster set $J_1$. The cluster computation in the other modes is performed in a similar way by the corresponding group of processes.

Each process holds a part of the tensor data assigned to it. 
For each slice $i$, we compute the top eigenvector of its covariance matrix and multiply it with the corresponding eigenvalue. Their product $\la_i\bv_i$ is the $i$-th column of the local matrix $M$. In practice, we compute them using the fast and well-known power iteration method. An alternative method to compute them with rapid convergence is the so-called Rayleigh quotient iteration method
\cite{trefethen1997numerical,kress2012numerical}. 
 
After this loop, each process holds a local maximum $(\la_0)_{local}$ between the top eigenvalues and a portion of the unnormalized global matrix $V$ in each process. We obtain the global maximum $\lambda_0$ using a reduction with result redistribution between all the processes of this group so that each process holds the global maximum eigenvalue. Similarly, the matrix $V$ is formed by concatenating the local portions of the $V$ matrix held by the processes of the group, and the result is redistributed between them. In practice, these operations are implemented in MPI by the collective operations {\tt MPI\_Allreduce} and {\tt MPI\_Allgather}.

Afterwards, the normalized form of the  matrix $M$ and $V$ are called $V\_local$ and $V$, respectively. We need to keep in mind that 
the  $i$-th column of $V$ contains the normalized product of the top eigenvalue and top eigenvector of the covariance matrix of the $i-$th slice of the tensor $\cT$ in the corresponding mode.

At this point of the computation, each process holds its local portion of the matrix $V$, and the $V$ matrix itself. Therefore, the matrices $V^TV$ can be computed in parallel. Each process computes the product  $V\_{local}^T V$ and takes the absolute values of all its entries. The result represents a subset of rows of the matrix ${\bf C}$. The number of columns of the matrix ${\bf C}$ equals the number of  slices in a corresponding mode of $\cT$. If we define $C\_local = V\_{local}^T V$, the concatenation along rows of $C\_{local}$ delivers the full matrix ${\bf C}$.

The sum of entries of each row of the matrix $C\_{local}$ are a subset of the vector $\mathbf{d}$ entries. Then, the concatenation of all entry subsets  of $\mathbf{d}$ held by the group processes is necessary for the cluster selection. We recall that the $i$-th entry of $\mathbf{d}$ is the marginal similarity of the $i$-th slice in the corresponding mode.

Then, the determination of each clustering set $J_1, J_2$, and $J_3$ is computed by one process of each group. We are calling this process the \emph{root process} in Algorithm \ref{algo:multi-slice-parallel}, and we can take an arbitrary process of the group (in practice, process 0 of the group).

In the last step, these group root processes gather the cluster sets $(J_1, J_2, J_3)$ on one processor (the global root process), which delivers the output cluster as a tuple $(J_1,J_2,J_3)$.

\section{Performance evaluation}
\label{Parall-experiments}

We evaluate the  performance of our algorithm
scheme on synthetic datasets that are, in fact, very similar to those of the MSC case \cite{andriantsiory2021multislice}. 
However, the size of the dataset will be $1000 \times 1000 \times 1000$, which is much larger than the experiments performed for MSC. In fact, it is 
\begin{equation*}
1000/50 = 20 \times \text{ larger than the sequential MSC  experiment}. 
\end{equation*}
Three sets $J_1, J_2$, and $J_3$ stand for the clusters from mode-1, mode-2 and mode-3 of the tensor, respectively. We assume that we have the same  number of slices in each mode, i.e. $m_1 =  m_2 = m_3$, and the number of elements inside the cluster is also equal ($|J_1|= |J_2| = |J_3| = l$). We construct the signal tensor  as follows:
    \begin{eqnarray}
\bv_i  = \left\{ 
\begin{array}{l l}
  1/\sqrt{l } &  \text{if } i\in J_1\\
  0 &  \text{otherwise}\\ \end{array} \right., 
  \bu_i  = \left\{
\begin{array}{l l}
  1/\sqrt{l } &  \text{if } i\in J_2\\
  0 & \text{otherwise}\\ \end{array} \right., \nonumber
\end{eqnarray} 

\begin{eqnarray}
  \bw_i  = \left\{ 
\begin{array}{l l}
  1/\sqrt{l} &  \text{if } i\in J_3\\
  0 & \text{otherwise}\\ \end{array} \right. \nonumber
\end{eqnarray}
Along the experiment, we fix the value of $l$ to be 10\% of the size of mode-$i$,  $l= \lfloor 10 m_i/100\rfloor$ for $i\in\{1,2,3\}$.  

The source code can be found at this link: 
https://github.com/ANDRIANTSIORY/MSC\_parallel. 
We now detail our experiments. 

\noindent
{\bf Implementation: }
We have run the performance evaluations on the Grid’5000 platform \cite{Grid5000}, using  \textbf{Gros} cluster in the site of Nancy. It is made of 124 nodes, each of which features to a CPU Intel Xeon Gold 5220, 18 cores/CPU, 96GB RAM, 447GB SSD, 894GB SSD, 2 x 25Gb Ethernet. The operating system deployed on the nodes is a
Debian 11 with a Linux kernel 5.10.0. All the code was compiled using g++ 10.2.1 with OpenMPI 4.1.0 and Python 3 using MPI bindings from mpi4py.

\paragraph{Evaluating MSC parallel algorithm depending on the signal weight}

This experiment is one of the most natural and simply checks that the
new version of MSC still keeps its  clustering performance and quality when applied on larger 3rd-order tensor datasets.  Such an experiment may be seen as a non-regression test. 
We generate a tensor dataset of dimension $1000\times 1000\times 1000$ and vary the weight $\gamma$ of the signal tensor from $100$ to $900$ with a step size equal to $50$. For each value of $\gamma$, we repeat the experiment 10 times,  re-sampling each time the noise. For each output of the algorithm, we evaluate the cluster quality by computing  the recovery rate (rec) and similarity index (sim): 
\begin{eqnarray}
   { \rm rec} =\frac{1}{3}\sum_{i=1}^{3} \frac{|J_i \cap \hat{J}_i|}{|J_i|} 
   \qquad 
    {\rm sim} = \frac{1}{3} \sum_{i=1}^{3}\;  \frac{1}{|\hat{J}_i|^2}\; \sum_{i,j \in \hat{J}_i } \; c_{ij} 
    \label{eq:par:sim}
\end{eqnarray}
Both measures lie 
between 0 and 1. 
As the name suggests it, the recovery rate measures the rate of output covering compared to the desired clusters.
Independent 
of the expected clusters, the similarity score 
determines if the output cluster is strongly similar or not. 
If  ${\rm sim}$ is close to 0 (resp. to 1), it means that the output is of bad (resp. good) clustering quality. We run this experiment on an MSC parallel version with 27 processes.
\begin{figure}[htbp]
    \centering
    \hspace{-0.2cm}
    \begin{subfigure}[b]{0.4\textwidth}
        \centering
        \includegraphics[scale=0.45]{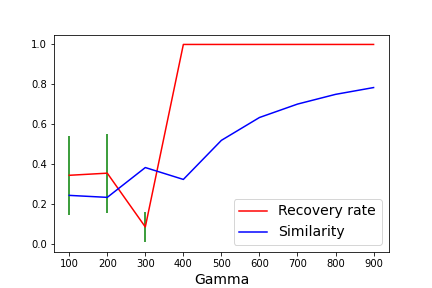}
        \caption{$\epsilon= 10^{-5}$}
    \end{subfigure}
    \begin{subfigure}[b]{0.4\textwidth}
        \centering
        \includegraphics[scale=0.45]{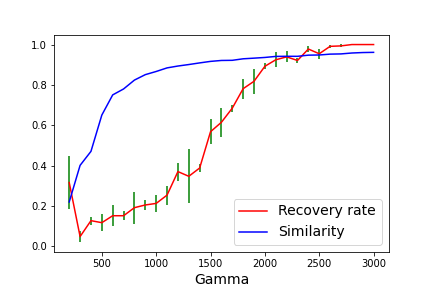}
        \caption{$\epsilon = 1.2 * 10^{-6}$}
    \end{subfigure}
	 	
\caption{Similarity and recovery rate mean  varying $\gamma$  for dataset of size $1000^3$. On the top $\epsilon$  violates the theorem \ref{thm1} hypothesis, on the bottom it fulfills it.}
\label{fig:par:findSigma}
\end{figure}

In figure \ref{fig:par:findSigma},  the blue line represents the similarity mean of the 10 experiments (standard deviation is negligible).
The red curve depicts the recovery rate (with standard deviation in green bars). 
In figure \ref{fig:par:findSigma}(a), the recovery rate shows that the cluster is found quickly but the cluster remains of relatively weak similarity ($<$80\%). Oppositely, in figure \ref{fig:par:findSigma}(b),  for an $\epsilon$ obeying the theorem condition,  the similarity shows very good cluster quality. Furthermore, in this case, we were able to link the recovery rate and similarity progression, replicating the same features as the original MSC algorithm. This means that our parallel version retained the main features required for a performant clustering analysis.

\paragraph{Strong scalability of the MSC parallel algorithm}

This experiment aims to display the  
 execution time 
of the MSC parallel version when the number of processes increases.
 We set the size of the datasets to  $1000\times 1000\times 1000$,  and  
also perform our computation for a sample of values of $\gamma$: 900, 1500 and 2000, 
 and $\epsilon= 1.2*10^{-6}$.
As explained earlier, the number of the processes needs to be a multiple of 3. We start with 6 processes and until 96 processes with stepsize 9. 
For each fixed number of processes, we evaluate the runtime of the execution 10 times (resampling the noise), then compute the mean and standard deviation  of these procedures. The result is shown in figure \ref{fig:distr_D1000_Pchange}.

\begin{figure}[htbp]
    \centering
        \includegraphics[scale=0.5]{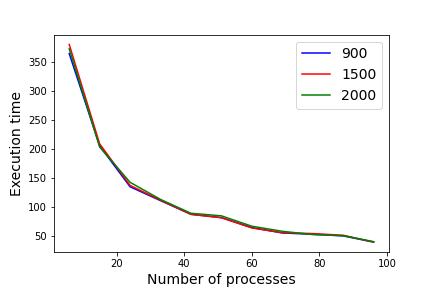}
	 \caption{Computational time with a different number of processes (6 to 96) for a dataset of size $1000\times 1000 \times 1000$ with three different value of the signal strength.}
	 	\label{fig:distr_D1000_Pchange}
\end{figure}

Figure \ref{fig:distr_D1000_Pchange} shows 
the computational cost for the MSC procedure itself. 
 The curve presents the mean of the  execution time of the 10 experiments for each fixed number of processes.
The maximum value of the standard deviation 
is insignificant. 
For the different fixed values of $\gamma \in \{900, 1500, 2000\}$, the curve always presents the  same features: (1) From  6 up to 96 processes, 
 the execution time drops drastically for an increasing number of computing processes; 
 (2) we see that the computational time of the MSC algorithm decreases abruptly with a large gap between 6 to 15 processes ($-150$ seconds).  
 Afterwards, the acceleration slows down for the next  steps. 
 In conclusion, this shows that our   parallel computing successfully fulfills its goals.

\paragraph{Execution time for a fixed number of processes and varying data sizes}



This experiment sets the number of processes and runs the code with different sizes of the tensor dataset. It evaluates the \emph{data} scalability. The tensor size varies from 200 to 1400 for the parallel version and for the sequential version, with a step size of 200. The cluster size is always 10\% of the dimension in the different modes. Since we are interested in the execution time and figure \ref{fig:distr_D1000_Pchange} shows that the computation time does not depend on $\gamma$, we set the value of $\gamma$  to the dimension of the first mode of the tensor dataset for all experiments. For each data size, we run the code 10 times and show the average value, recording the time computation for the MSC  parallel versions with a given number of processes (33, 60, 87 and 123) and the sequential version.  
The output of these experiments is presented in figure \ref{fig:parallel_15_123}.

\begin{figure}[htbp]
        \centering  
        \hspace{-0.2cm}
        \begin{subfigure}[b]{0.4\textwidth}
           \centering
           \includegraphics[scale=0.45]{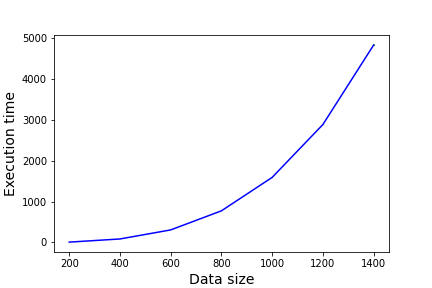}
           \caption{Sequential version.}
        \end{subfigure}
        \begin{subfigure}[b]{0.4\textwidth}
            \centering 
            \includegraphics[scale=0.45]{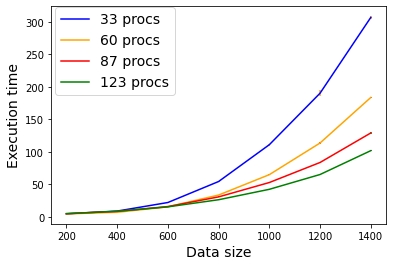}
            \caption{Parallel version.}
        \end{subfigure}
\caption{Computation of the MSC.}
 \label{fig:parallel_15_123}
\end{figure}

In these pictures, 
the horizontal axis collects the datasizes and the vertical axis presents the execution time measured in seconds. 
First, the curves 
show that the necessary time to complete the clustering computation increases with the dataset size, as expected. 
  However, we must pay attention to the growth rate of the running time of these different MSC versions. 
In the final datasize $1400$, we see that the execution time decreases if we increase the number of processes. The fastest computation is for 123 processes, it takes $102s$(1min42s)  to run MSC to the dataset of size $1400^3$, see figure \ref{fig:parallel_15_123}. 
Meanwhile, 
for the same size of a tensor dataset, the computation time with the sequential version
stretches to a bit below $5000s$ (1h23min).

Based on these results, figure \ref{fig:speedup} shows the speedup between the sequential execution time and the parallel execution time for different numbers of processes. Generally, the speedup becomes significant  with larger datasize, and the highest speed corresponds to parallel MSC with the larger number of processes. Our experiment is conclusive, there is a radical time gain  using the MSC parallel version
 (e.g.  for a datasize 1400, 
 parallel MSC with 123 processes is 48 times better than the sequential MSC).

 \begin{figure}[h!]
        \centering
        \includegraphics[scale=0.5]{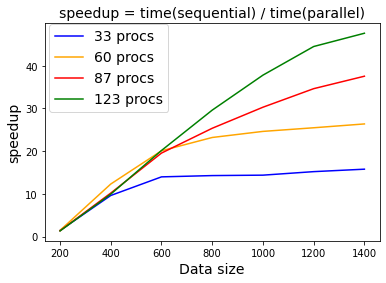}
        \caption{Parallel speedup.}
        \label{fig:speedup}
\end{figure}

\section{Communication}

Lastly, we present where time is spent during the communication steps between the different processes in figure \ref{fig:profile}. To do so, we choose to use TAU (Tuning and Analysis Utilities) and compute the evaluation for 33, 60, 87 and 123 processes with tensor data of size $1000 \times 1000 \times 1000$. In the figure, we only plot the maximum time consumption for each MPI method. 

For a given data size, our evaluation reveals
that the time spent in {\tt MPI\_Allreduce} operations decreases when more processes are used.  Since the data on each process is smaller, smaller messages are sent between processes and bandwidth-bound communications are faster. The time spent in {\tt MPI\_Gather} and {\tt MPI\_Gatherv} increases with 123 processes, suggesting that on this number of processes, the data is small enough for the operation not to be bandwidth-bound anymore. 
Secondly, less time is spent in the {\tt MPI\_Gather} operations than in the {\tt MPI\_Allreduce} operation.

\begin{figure}[htbp]
        \centering  
        \hspace{-0.2cm}
        \begin{subfigure}[b]{0.4\textwidth}
           \centering
           \includegraphics[scale=0.5]{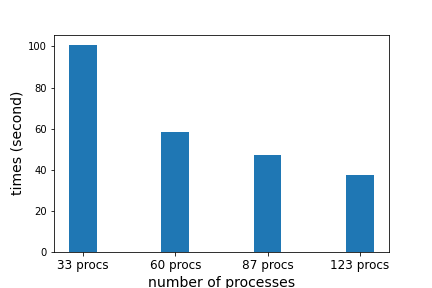}
           \caption{Time spent in {\tt MPI\_Allreduce}.}
        \end{subfigure}
        \begin{subfigure}[b]{0.4\textwidth}
            \centering 
            \includegraphics[scale=0.5]{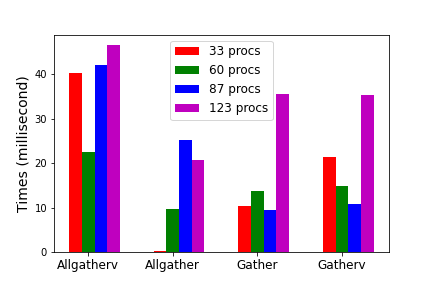}
            \caption{Time spent in {\tt MPI\_Allgather}.}
        \end{subfigure}
\caption{Time consumed for the data transfer inside algorithm \ref{algo:multi-slice-parallel} with data of size $1000\times 1000 \times 1000$. }
 \label{fig:profile}
\end{figure}

\section{Conclusion}
We have introduced a parallel algorithm for shared memory for a recent clustering algorithm called MSC. We showed that the  MSC parallel version outperforms the sequential one as far as performance is concerned. Moreover, the MSC parallel version ensures: (1) preservation of the cluster quality, (2) performance gain, and (3) scalability.

Our parallel MSC algorithm relies on two properties of the MSC algorithm. The first one is based on the fact that MSC computes the clustering of one dimension after the other and independently from each other. The second idea is the fact that the MSC algorithm deals with multiple and independent spectral analyses. It contains subroutines for constructing piece-by-piece algebraic objects (vectors and matrices), each piece of information  develops from part of the data. Such features make the MSC algorithm particularly suitable for parallelization. These properties partially adapt to other algorithms. Indeed, the so-called Tensor Biclustering and Triclustering \cite{FeiziNIPS2017} share a feature with MSC:  these algorithms also perform clustering analysis one dimension at a time. Extending this work to other algorithms for higher dimensional datasets is certainly worth investigating in the future. 

Our algorithm relies on a distributed data layout. The data is either distributed or produced on the processes themselves. It would be interesting to integrate it in distributed applications that generate the data on these processes and avoid moving the data between steps of their computation. The fact that it never requires all the data to be on a single process makes it possible to execute it on  large datasets that do not fit in the memory of a single node. We can also investigate other parallel computing schemes. We can choose a Master-Worker scheme and see if a centralized aggregation of the results would be promising, rather than using collective operations. Finally, real-life experiments on the clustering of large-size tensor datasets ought to be addressed using MSC parallelization and verify that our algorithms give valid and proper results. These aspects are also left for future investigation. 


\section*{Acknowledgements} 
Experiments presented in this paper were carried out using the Grid'5000 testbed, supported by a scientific interest group hosted by Inria and including CNRS, RENATER and several Universities as well as other organizations (see https://www.grid5000.fr).

\bibliographystyle{unsrtnat}
\bibliography{biblio}

\end{document}